\documentclass[aps,prl,a4paper,10pt,twocolumn,superscriptaddress]{revtex4-1}
\usepackage[utf8x]{inputenc}
\usepackage{graphicx}
\usepackage{amsmath}
\usepackage[australian]{babel}
\usepackage{color}
\definecolor{dblue}{rgb}{0,0,0.6}
\usepackage[pdftex,plainpages=false,bookmarks=true,colorlinks,breaklinks,linkcolor=dblue,citecolor=dblue,urlcolor=dblue,filecolor=dblue]{hyperref}

\DeclareMathOperator{\erfc}{erfc}

\usepackage{color}

\begin{document}

\title{Total Current Blockade in an Ultra-Cold Dipolar Quantum Wire}

\author{L.H. Kristinsd\'ottir}
\author{O. Karlstr\"om}
\author{J. Bjerlin}
\author{J.C. Cremon}
\affiliation{Mathematical Physics, LTH, Lund University, Box 118, 22100 Lund, Sweden}
\author{P. Schlagheck}
\affiliation{D\'epartement de Physique, Universit\'e de Li\`ege, 4000 Li\`ege, Belgium}
\author{A. Wacker}
\author{S.M. Reimann}
\affiliation{Mathematical Physics, LTH, Lund University, Box 118, 22100 Lund, Sweden}

\begin{abstract}
 Cold atom systems offer a great potential for the future design of new
 mesoscopic quantum systems with properties that are fundamentally different
 from semiconductor nanostructures. 
 Here, we investigate the {\it quantum-gas analogue} of a
 quantum wire, and find a new scenario for the quantum transport: Attractive
 interactions may lead to a complete suppression of current in the low-bias
 range, a \textit{total current blockade}. We demonstrate this effect for the
 example of ultra-cold quantum gases with dipolar interactions.
\end{abstract}

\maketitle

The electronic Coulomb blockade in mesoscopic quantum dots has been an intensive topic of research over the last two decades. The flow of electrons through a quantum dot  between two reservoirs turned out to be an extremely versatile tool for addressing a wide range of fundamental effects. Examples range from investigating the structure of electronic many-particle states~\cite{ReimannRMP2002,HansonRMP2007} and Kondo physics~\cite{GlazmanJETP1988,NgPRL1988,GoldhaberNature1998}, to quantifying the spin dephasing due to coupling to nuclear degrees of freedom~\cite{PettaScience2005,KoppensScience2005,KoppensNature2006}, or coherent effects~\cite{NilssonPRL2010}.

Ultra-cold atoms in traps are very similar to quantum dots -- a few quantum
particles confined by a (often low-dimensional and harmonic) potential. What
makes these systems particularly interesting is, that one essentially can
freely engineer their properties, and even control the shape and strength of
the inter-particle interactions. More recently this sparked great interest in
making (quantum-)logical devices with 
ultra-cold atoms and molecules analogous to those in electronics and
spintronics~\cite{Seaman2007,Pepino2009,Pepino2010,Qian2011,Bruderer2012,brantut2012}.

``Interaction blockade'' as the cold-atom analog of electronic Coulomb
blockade~\cite{Capelle2007} was experimentally first seen in tunneling
processes in optical lattices~\cite{Cheinet2008} and analyzed theoretically
for one-dimensional triple-well systems~\cite{Schlagheck2010}. 
Atom trapping with numbers down to single-atom precision was 
reported in a remarkable recent experiment by 
Serwane {\it et al.}~\cite{Serwane2011}, reaching  
the few-body limit with full control over confinement and inter-particle 
interactions. 
The experimental realization of {\it quantum transport} of cold atoms through a
small quantum few-body system that is brought in contact with two large atomic
reservoirs, however, has up to now posed a great experimental challenge. 
A first experimental breakthrough was reported recently in an  
experiment by Brantut {\it et al.}~\cite{brantut2012}, that clearly demonstrates
the possibility to engineer both a ballistic and a diffusive channel between
two cold atom reservoirs, opening up a host of new perspectives in mesoscopic
quantum physics. 

Inspired by this recent experimental progress, we study in this Letter  
the quantum
transport through wire-like confinement of a few ultra-cold fermions.
In the framework of the experiment by Brantut
{\it et al.}~\cite{brantut2012}, such a structure could be realized by two
optical barriers within the channel, created by focusing two
blue-detuned laser beams perpendicularly onto the channel. 

A particularly interesting aspect of studying transport with cold atoms or
molecules is the tunability of the interactions between the particles - often
being of contact type, and experimentally controlled by Feshbach resonances. 
Here, we choose to study fermions with electric dipolar interactions 
which can be
controlled by an external field~\cite{lahaye2009}. 
Changing the interactions from repulsive to attractive, 
we report the occurrence of \textit{total current blockade}, where 
the attractive interaction hinders transport for finite biases 
independent of the gate potential. 
While the total current blockade would also occur with attractive 
contact interactions, dipolar interactions also make it possible to  
study localization effects due to the long-range nature of the force, in
much analogy to electrons in quantum wires~\cite{KristinsdottirPRB2011}.  
\begin{figure}
 \centering
 \includegraphics[width=\columnwidth]{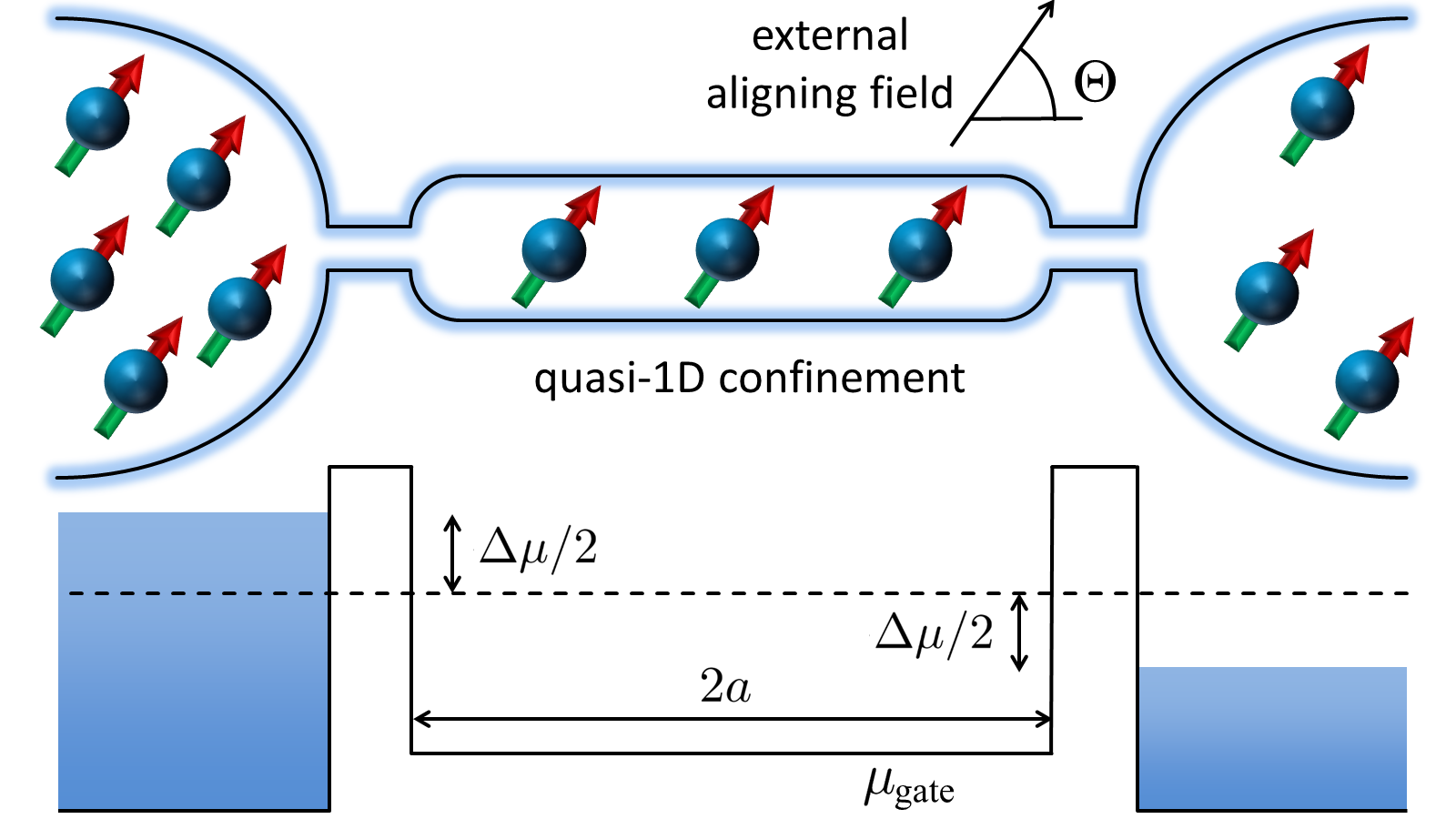}
 \caption{ 
  Upper panel: Schematic figure of the system. Lower panel: Sketch of the setup in analogy to the case of mesoscopic conductors. Two reservoirs  with a degenerate gas of ultra-cold spin-half dipolar particles are connected via a quasi one-dimensional structure, a ``wire'' of length $2a$. The difference in chemical potential between the reservoirs, $\Delta\mu$, creates a particle current if the dipoles can be added and removed from the wire. Levels in the wire may be tuned by a gate potential, $\mu_\text{gate}$. The interaction between the particles in the wire can be varied by the tilt angle of the dipoles, $\Theta$, and allows to observe significantly different current patterns.
 }
\label{f:schem_model}
\end{figure}
%

The setup of the system described above is 
sketched in Fig.~\ref{f:schem_model}. 
Similar to the recent study by Brantut {\it et al.}~\cite{brantut2012}, 
two fermionic reservoirs with controllable difference in chemical potential
$\Delta\mu$ are connected by a quasi one-dimensional trap. 
In this region, the potential energy of the
particles can be varied by the parameter $\mu_\text{gate}$ in full
analogy to electrons in gated semiconductor nanostructures.
The electric dipole moment $p$ of the particles can
be orientated along an external field by a tilt angle $\Theta$ with
respect to the $z$ axis along the quasi one-dimensional channel (see
Fig.~\ref{f:schem_model}).  One can thereby also minimize the 
dipolar component of the particle interactions in the leads, and 
stabilize the dipolar gas against collapse in the left and right
reservoirs, required to be 
two-dimensional and appropriately oriented with respect to
the external field. A small local variation of the orientation angle $\Theta $ 
allows inducing attractive or repulsive interactions locally within the
wire. 

{\em Model}.---%
The interaction between two dipoles with distance 
$r$ and angle $\theta_{rd}$ between the dipole orientation and particle 
separation is generally given by\cite{JacksonBook1998,Skinner1989}
\begin{equation}
 V_\text{dd} =
 \frac{d^2}{r^3}\left.\left(1-3\cos^2\theta_{rd}\right)\right|_{r>0}
 - \frac{4\pi}{3}Cd^2\delta^3(\mathbf{r})
\end{equation} 
The coupling
strength is $d^2=p^2/(4\pi\epsilon_0)$, where $p$ is the dipole moment
strength, $\epsilon_0$ is the vacuum permittivity and $C=1$. 
(For magnetic dipoles, 
$d^2=\mu_0\mu ^2/(4\pi)$ is significantly smaller, 
where $\mu_0$ is the vacuum
permeability, $\mu$ the magnetic moment and
$C=-2$.) While the first term provides the common angular dependence of
dipole-dipole interaction, the second term provides a contact
interaction which is frequently disregarded. Within the quantum wire, 
the dipoles are confined in $x$ and $y$ by a two-dimensional harmonic
oscillator of characteristic length $l_\perp$, rendering a quasi
one-dimensional system in the $z$-direction for small
$l_\perp$. Integrating over the lateral  $x$ and $y$ degrees of
freedom one arrives at an effective one-dimensional dipole-dipole
interaction
\begin{equation}
 V_{\text{dd}}^\text{eff}(z_1,z_2) = U_\text{dd}(\Theta) 
f\left(\frac{|z_1-z_2|}{l_\perp}\right) 
+\frac{2Cd^2}{3l_\perp^2}\delta(z_1-z_2)
\label{EqInteraction}
\end{equation}
with 
$f(u) = -2u + \sqrt{2\pi}(1+u^2)e^{u^2/2}\erfc(u/\sqrt2)$
where $\erfc$ is the complementary error function 
\cite{Deuretzbacher2010}.
The interaction coefficient
\begin{equation}
 U_\text{dd} = -\frac{d^2[1+3\cos(2\Theta)]}{8l_\perp^3}\,,
\end{equation}
can be either positive or negative depending on the dipole tilt angle
$\Theta$. If the dipoles are aligned in the $z$ direction
($\Theta=0^\circ$) they attract each other, $U_\text{dd}<0$, while
they repel one another, $U_\text{dd}>0$, if they are orientated
perpendicular to the $z$ direction ($\Theta=90^\circ$). For an
intermediate angle  ($\Theta_\text{crit}\simeq54.7^\circ$) this
long-range part of the dipole interaction vanishes. 

In the $z$-direction the wire is modeled as a finite square well (see
Fig.~\ref{f:schem_model}) of width $2a$ and barrier
height $V_0$. Applying the single-particle basis of eigenstates for
this potential well, the configuration interaction method (exact
diagonalization) is used to find the lowest energy states of $N=1$ to
$N=6$ dipoles in the quantum wire. Here the dipolar particles are
assumed to be spin-half fermions. In the following we use
$d^2=\hbar^2a/m$,  $l_\perp=0.14a$, and $V_0=300\hbar^2/ma^2$. 
(For RbK molecules with $p=0.57$~Debye \cite{NiNature2010} 
this corresponds to $a\approx 0.6\mu$m and 
an energy unit of $\hbar^2/ma^2\approx k_B 10{\rm nK}$)

Transitions between states of different $N$ 
occur due to particle exchange with the
reservoirs, as described by 
rates $\Gamma_{a\rightarrow b}$  evaluated by Fermi's
golden rule. The corresponding matrix elements between the 
many particle state are evaluated  following the work of
\cite{PedersenPRB2005,CavaliereNJP2009} 
for mesoscopic
electronic systems. Assuming, that the occupations of the single-particle states in the 
particle reservoirs are given by Fermi-functions with 
$k_BT=0.02\hbar^2/ma^2$,
this provides a Pauli master equation for the
probabilities of the different many-particle states in the confinement region.
For the stationary state we obtain the
(particle) current $\dot{N}$ between the reservoirs and the
(differential) conductance $G=\mathrm{d}\dot{N}/\mathrm{d}\Delta\mu$.

\begin{figure}
 \includegraphics[width=0.97\linewidth]{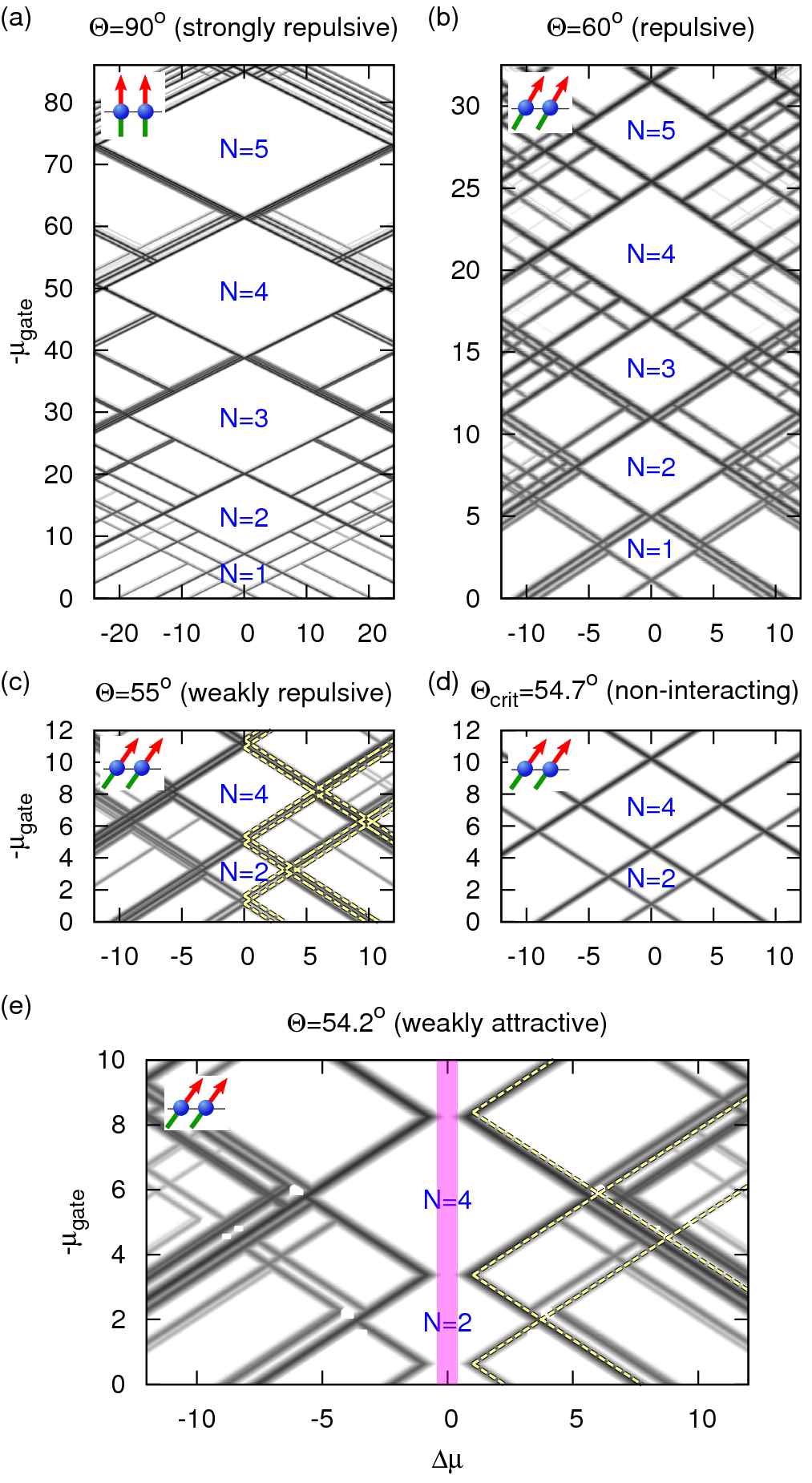}
 \caption{ 
  Conductance between the particle reservoirs as a function of reservoir
  potential difference $\Delta\mu$ and gate potential $\mu_\text{gate}$. Here
  the contact part of the dipolar interaction is neglected and the long-range
  part, which can be tuned by the angle $\Theta$ of the external field,
  changes from (a) strongly repulsive, via (d) non-interacting, to (e) 
  the weakly
  attractive case. The region of total current blockade for attractive
  interaction is colored in magenta in (e). The dashed (yellow) lines in (c) and (e)
  indicate the results of a simplified quasi-independent-particle model. 
The calculations were performed for $d^2= 1.0~{\hbar ^2 a/m} $ and 
$l_{\perp } = 0.14 a$. The $\mu$-scale is in units of $\hbar^2/ma^2$. Note the different scales in panel (a) and (e).
 }
 \label{f:diamonds}
\end{figure}

%
\textit{Main results}.---%
First we neglect the contact term of the dipolar interaction, which can be
eliminated by Feshbach resonances~\cite{werner2005}, and obtain the
conductance diagrams displayed in Fig.~\ref{f:diamonds}. For repulsive
interaction between the dipoles, see Figs.~\ref{f:diamonds}(a)-(b), the
conductance diagrams resemble those of Coulomb blockade as intensively 
studied by electron 
transport in mesoscopic structures \cite{ReimannRMP2002}.  

In the diamond-shaped regions of vanishing conductance,  
the particle number $N$ in the wire is fixed, and the current between the
reservoirs is strongly suppressed. 
At the borders between the $N$ and $(N+1)$-particle region, conductance is
possible due to single-particle transitions. 
This scenario does not depend on the specific form of the 
(repulsive) interaction~\cite{Capelle2007,Cheinet2008}.  

In contrast to electronic systems, however,
the tunability of the interaction for dipolar fermions allows to reduce the
interaction strength (Figs.~\ref{f:diamonds}(c)-(d)) and even reach a scenario where the interactions become attractive, see Fig.~\ref{f:diamonds}(e):
We then obtain a total current blockade  at low detuning $\Delta\mu$ independent of the gate potential $\mu_ \text{gate}$. This is a clear-cut signature of attractive interactions.

The total current blockade is associated with the vanishing of the diamonds
for odd $N$. This can be understood by the two-fold degeneracy of the single
particle levels due to the particle spin: The first particle enters the system
at the level energy, while the second particle experiences an additional
interaction energy $U$ between the spin-degenerate particles in a level. The
single occupancy of the level, i.e. a state with odd $N$, is stable if the
reservoirs allow for adding the first particle, but not the second. For $U>0$
(i.e. the conventional repulsive Coulomb interaction or $\Theta > 54.7^\circ$
for the dipoles studied here) this is possible.  Thus, the blockade diamonds
with an odd number of particles $N$ and lines of finite conductance at the
separation to the blockade diamonds with even $N$ appear in
Figs.~\ref{f:diamonds}(a),(b). With decreasing interaction the width of all
diamonds shrinks and the width of the odd-$N$ diamonds vanishes at $U=0$ as
can be seen in Fig.~\ref{f:diamonds}(d). Now, for negative $U$ the situation
of a single fermion in a level is unstable as it attracts a particle with the
opposite spin. This instability does not allow for configurations with odd $N$
for low $\Delta \mu$. Therefore, single-particle transitions between the
reservoir and the wire are excluded, resulting in the absence of current flow
in the region of total current blockade, see the magenta shaded area in
Fig.~\ref{f:diamonds}(e). (The case of two-particle transitions is addressed
below.)

For weak interactions, this can be quantified 
by a quasi-independent-particle model: The single-particle  
level energies of the quantum well are approximated by $n^2E_1$
where $n=1,2,\ldots$ and $E_1$ is the single-particle ground state energy. 
Using the analytic eigenfunction of the infinite
well, we approximate the interaction energy by first order
perturbation theory. Then the energy difference between the $N+1$ and
the $N$-particle ground state is given by
\begin{equation}
 \mu_{N+1} = \mu_\text{gate} + (n+1)^2 E_1 + \left(\tfrac{2}{3}n+\delta\right) U
\end{equation}
where $U\equiv 3 l_\perp U_{\rm dd} / a$ and $\mu_\text{gate}$ is
the gate potential relative to the bottom of the well. Here, $n=
N/2$ and $\delta = 0$ for even $N$ while $n=
(N-1)/2$ and $\delta = 1$ for
odd $N$. The lines of the diamonds are given by the crossing
points of $\mu_{N+1}$ with the chemical potential $\pm\Delta\mu/2$ in
the left or right reservoir, respectively.
The corresponding dashed (yellow) lines shown in Figs.~\ref{f:diamonds}(c,e) agree well with the main
conductance lines obtained from the full many-particle
calculation. Thus, correlations do not play any essential role here.
\begin{figure}
 \includegraphics[width=\linewidth]{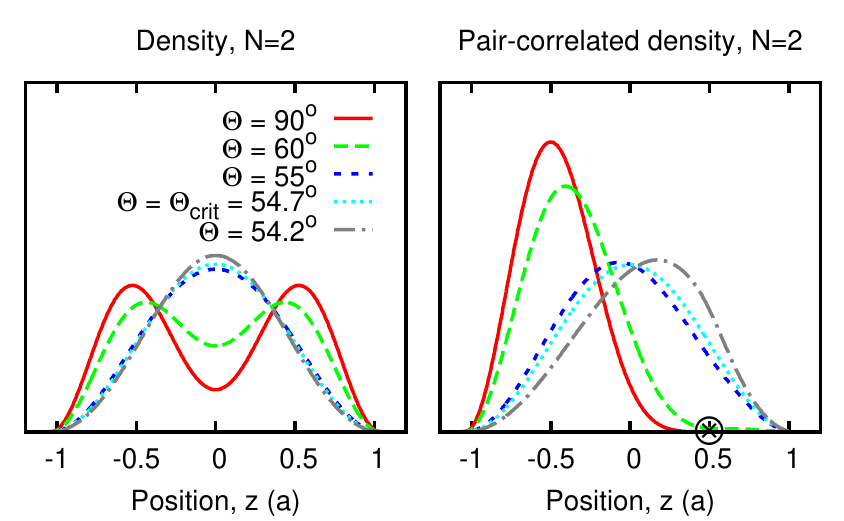}
 \caption{ 
  Particle density (left) and pair-correlation function (right) for $N=2$ particles at the tilt angles $\Theta$ used in Fig.~\ref{f:diamonds}(a)-(e). For the pair-correlation function one particle is fixed at the position marked with the symbol $\boldsymbol{\otimes}$. As the interaction goes from strongly repulsive ($\Theta=90^\circ$) to weakly attractive ($\Theta=54.2^\circ$) the two particles evolve from a localized state to a delocalized state with a slight tendency to clustering.
 }
 \label{f:densities}
\end{figure}

In contrast, such an approach does not hold for stronger interaction strengths. Here the many-particle states show strong localization effects as shown in Fig.~\ref{f:densities} for the two-particle states. For $\Theta=90^{\circ}$ and to a smaller extent for $\Theta=60^{\circ}$, one observes two peaks in the particle density (left panel), and the pair-correlation function (right panel) shows that the probability to find the two particles within the same peak is strongly reduced. This is the scenario of Wigner localization as very recently studied theoretically for cold polar molecules in \cite{knap2012}. In full analogy to mesoscopic electron conduction \cite{KristinsdottirPRB2011}, signatures of this localization can be clearly detected in the conductance plots Fig.~\ref{f:diamonds}(a)-(b) where several, almost degenerate lines are observed on the top of the diamonds, which result from spin excitations of the localized particles.

We note that this scenario of Wigner localization would not arise in transport
processes with atomic species that only interact through a  contact
interaction potential. 

\textit{Improved interaction model}.---%
For attractive interaction ($\Theta=54.2^\circ$), the pair-correlation
function is shifted to the right, see the right panel of
Fig.~\ref{f:densities}, \textit{i.e.} the probability to find both fermions on
the same spot is enhanced for the groundstate. In this case the contact
interaction in Eq.~(\ref{EqInteraction}) becomes relevant. Taking this term
into account provides some modifications of the scenario depicted in
Fig.~\ref{f:diamonds}, while the main features remain. For the case of
electric dipoles, the contact interaction is repulsive and compensates a part of the long-range attraction, so that smaller angles $\Theta$ are required to observe the vanishing of the diamonds with odd $N$. Furthermore, since the particle density  increases with the number of particles $N$, the contact interaction becomes more relevant for higher $N$, and thus smaller angles are required for the vanishing of diamonds with higher $N$. We have observed this for e.g. $\Theta=46^\circ$, where the $N=1$ diamond has already vanished, while the $N=3$ diamond is very small and the  $N=5$ diamond is still well established. In this case the total current blockade due to the attractive interaction extends only over a part of the spectrum.
\begin{figure}
 \includegraphics[width=\columnwidth]{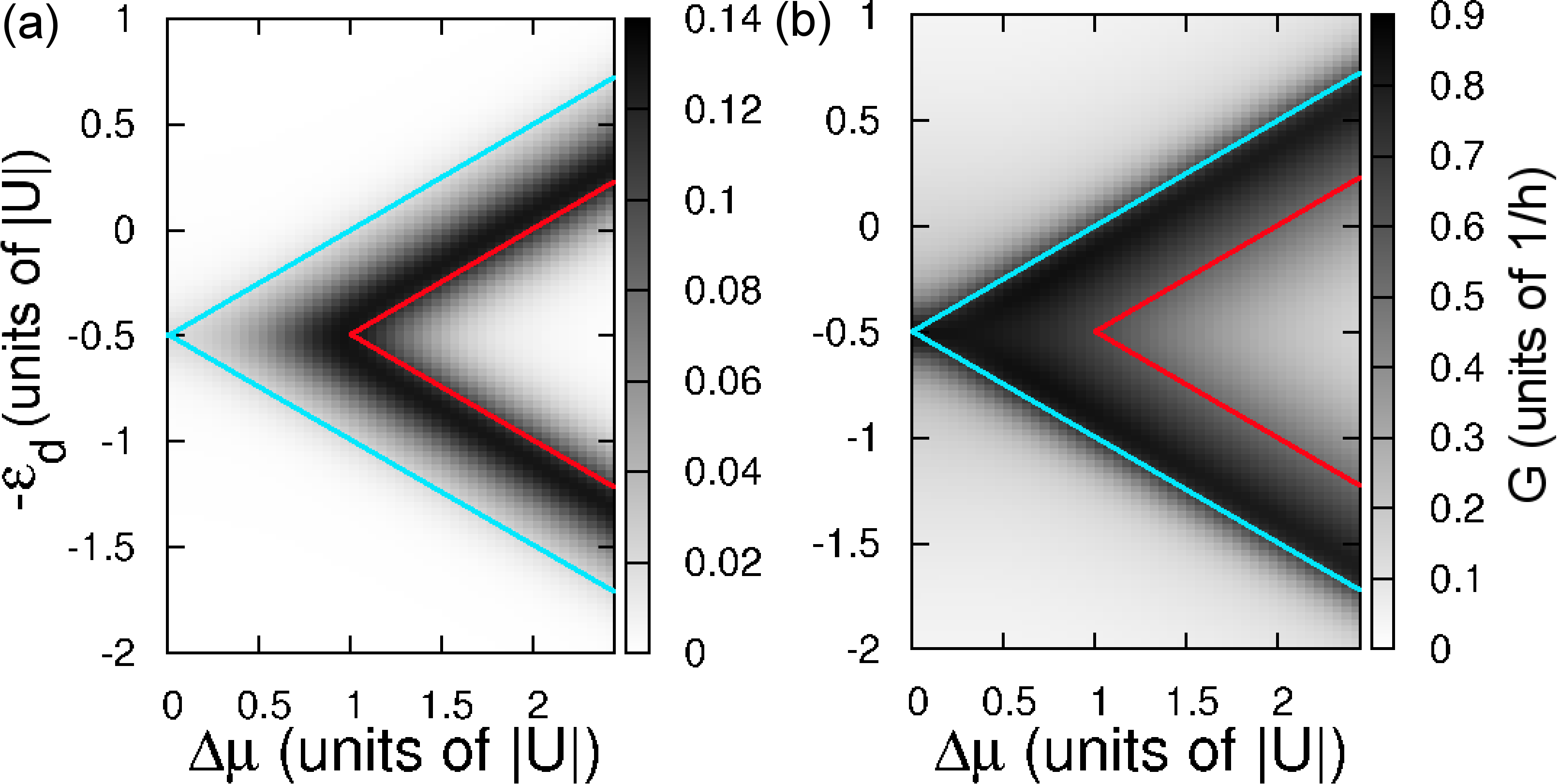}
 \caption{ 
  Conductance through a single spin-degenerate level at energy $\varepsilon_d$
  for the case of inter-particle attraction, resulting in negative charging
  energy $U<0$. The temperature is $k_BT=|U|/10$, and the couplings are 
  $\Gamma_L=\Gamma_R=|U|/50$ for (a), and $\Gamma_L=\Gamma_R=|U|/4$ for (b).
The red and blue lines show the onset of sequential and pair-tunneling, respectively.
 }
\label{f:pairtunneling}
\end{figure}
%
\textit{Pair-tunneling}.---%
As discussed above, the  situation of a single fermion in a level is unstable for the case of attracting particles, $U<0$. Thus, single-particle transitions between the reservoir and the wire are excluded for sufficiently low vales of temperature and bias $\Delta\mu$. Here we want to illuminate the role of two-particle transitions, which may occur due to higher-order processes in the coupling between the reservoirs and the wire~\cite{Koch2006,Lopez2007}.

There are two kinds of processes: Normal co-tunneling, which results in a weak
background conductance for any bias, and pair-tunneling which, neglecting the
effects of temperature and lifetime broadening, is only allowed for
$|E_{2n+2}-E_{2n}|<\Delta\mu $, where $E_N$ is the ground state energy of the
$N$-particle state and $n$ is an integer. When present, pair-tunneling gives a
more pronounced contribution than co-tunneling~\cite{Koch2006}. Being of
second order, these processes scale as $\Gamma^2$, where $\Gamma$ is the
single particle transition rate. Thus, for sufficiently weak couplings they can be neglected compared to sequential single-particle tunneling.

Figure~\ref{f:pairtunneling} shows the differential conductance of a single spin-degenerate level with $U<0$, calculated by the second order von Neumann formalism \cite{PedersenPRB2005,PedersenPHE2010}.  For weak contact coupling, Fig.~\ref{f:pairtunneling}(a) displays only a small conductance at low values of $\Delta\mu$. This can be attributed to a weak pair-tunneling background and to the temperature broadening $\sim\!3k_BT$  of the direct tunneling peaks at $\varepsilon_d=-U/2\pm (\Delta\mu+U)/2$ for $\Delta\mu>-U$, which correspond to the red lines in Fig.~\ref{f:pairtunneling}. This demonstrates that the total blockade of conductance is verified for $\Gamma\ll k_BT\ll |U|$, as is the case in Fig.~\ref{f:diamonds}(e).

On the other hand, one has to keep in mind that, as $\Gamma$ approaches $U$,
pair-tunneling becomes energetically allowed. Hence, we observe the onset of
conduction along the blue lines $|E_2-E_0|=\Delta\mu$ in
Fig.~\ref{f:pairtunneling}(b). (In our case  $E_2=2\varepsilon_d+U$ and
$E_0=0$.) Normal co-tunneling can also be observed as a weak background present
at all $\Delta\mu$ and $\varepsilon_d$. Thus, the total blockade of
conductance does not persist at strong couplings between the wire and the
reservoirs. For even higher couplings, our model fails and Kondo-like effects
become important \cite{KochPRB2007}. This shows that the
energy barriers confining the wire cannot be arbitrarily weak for 
the observation of the total current blockade, as otherwise pair-tunneling masks the scenario.

\textit{Experimental challenges}.---%
From the experimental point of view, measuring a weak atomic current in a
mesoscopic transport process appears challenging. In the experimental studies
of quantum transport through atom traps by Brantut {\it et al.}~\cite{brantut2012}, the integrated current is measured by a sensitive detection of population differences in the reservoirs. This opens up a new field of mesoscopic physics research. Complementary experimental information on the atomic current could, for instance, be inferred from a time-of-flight absorption image that renders the momentum distribution of the transported atoms. As an alternative, a stimulated Raman adiabatic passage (STIRAP) of the atoms could be induced by irradiating the transport region with two spatially displaced laser beams (see, e.g., Ref.~\cite{Kuhn2002}). An atom that propagates through this irradiated region would then necessarily transfer a photon from one of the laser beams to the other, while an atom that propagates in the opposite direction would revert this photon transfer. A careful measurement of the net photon transfer between the beams after a suitable evolution time would then give rise to the integrated atomic net current across the atom-photon interaction region. We remark that standard techniques to detect individual atoms using fluorescence imaging \cite{Bakr2010,Sherson2010} or electron beams \cite{Wuertz2009} would not work in this context as they do not distinguish between left-moving and right-moving atoms.

\textit{Conclusions}.---%
We have shown that dipolar quantum gases allow for the observation of a total current blockade for small differences in chemical potentials between the reservoirs. In this context the often neglected contact interaction part of the dipole-dipole interaction turns out to repress the onset of total current blockade.

From the experimental side, studies of quantum transport with ultra-cold atoms and the many-body effects of interaction blockade are still in their infancy. Here, we highlighted the prospects for the specific example of a few-body system with dipolar interactions between the confined atoms. We demonstrated the possibilities offered by the tunability of the dipole-dipole interaction in a quasi one-dimensional geometry by an external field.

We thank the Swedish Research Council and the Nanometer Structure Consortium at Lund University (nmC@LU) for financial support. Furthermore, we thank Georg Bruun, Frank Deuretzbacher, Georgios Kavoulakis and Chris Pethick for discussions regarding the dipolar interaction potential.

\end{document}